\documentstyle[11pt,newpasp,twoside,epsf]{article}
\begin{document}
\title{Discovery, identification, and study of variability
in gamma-ray point sources}
\author{M. Fiorucci$^{1, 2}$, N. Marchili$^1$, G. Tosti$^{1, 2}$, S. Ciprini$^{1, 2}$, 
F. Marcucci$^{1, 2}$, A. Tramacere$^1$}
\affil{(1) Dipartimento di Fisica, Universit\`a di Perugia, 06123 Perugia, Italy
(2) I.N.F.N. sez. di Perugia, Via Pascoli, 06123 Perugia, Italy}

\begin{abstract}
We present preliminary results of a statistical analysis obtained with a sample
of blazars observed at the Perugia Astronomical Observatory since 1992.
We briefly show how these statistical results can be useful to discriminate
faint variable sources against the background noise. This technique, together
with the more traditional ones, may be used to discover and identify high-energy
point sources.
\end{abstract}

\section{Statistical analysis of variability}

Many of the expected high-energy sources are variable, like blazars
which emit signals that often appear to vary chaotically with time.

Does such signal result from the superposition of many random events
(randomly generated shocks in outflowing gas; randomly appearing
reconnections of magnetic field lines; etc.)?
Or they are governed by some global physical mechanism described by
non-linear differential equations, which shows deterministic chaos
(coherent variability of clouds of electrons; instability of a shock
at the base of the jet forming region; etc.)?
Finally, is it possible to find same kind of periodicity, at least
in particular intervals of time
(hot spots in the rotating accretion disk; helicoidal movements of the
jet; etc.)?

For a reasonable answer to these questions, we can use - for example -
one or more of the following statistical analysis:

\begin{itemize}

\item{\it correlation, autocorrelation, power spectrum estimation, with FFT:}
future high energy missions working in scanning mode (like GLAST) probably will 
provide data well sampled at evenly spaced intervals, that can be processed with 
the Fourier Transforms;

\item{\it correlation, autocorrelation, power spectrum estimation, with
the Lomb-Scargle periodogram}, (Lomb 1976, Scargle 1982) for irregularly or
incompletely sampled data;

\item{\it correlation and autocorrelation with the Discrete Correlation
Function (DCF)} described by Edelson \& Krolik (1988), useful for unevenly 
spaced data;

\item{\it first order Structure Function (SF)}, because there is a simple 
correspondence between power laws in the frequency space Fourier analysis,
and the SF analysis in the time domain (see, e.g., Hufnagel \& Bregman (1992);

\item{\it detrended fluctuation analysis}, that permits the detection of
intrinsic self-similarity embedded in a seemingly nonstationary time series
(see, e.g., Peng et al. 1995);

\item{\it the wavelet methods}, that together with principal component analysis
and filtering can be used to extract the ``deterministic'' components from the
time-series (see, e.g., Liszka \& Holmstr\"om 1999);

\item{\it a search for chaotic behavior}: the most popular measure of an
attractor's dimension is the correlation dimension, first defined by
Grassberger \& Procaccia (1983).

\end{itemize}

We are using these and other less-known statistical techniques with the 
BVR$_c$I$_c$
database of the Automatic Imaging Telescope at the Perugia astronomical
Observatory (Tosti et al. 1996). Preliminary results have been already
described in Fiorucci et al. (1999, 2000), and suggest
that the observed variability is truly stochastic and is not caused by
deterministic chaos. Moreover, many of these
statistical descriptors lead to the provisional conclusion that variability
of blazars is self-similar on
time-scales from a few days up to some years, and it is characterized by
power-law power spectrum $PSD(f) \propto f^{-\alpha}$, where the spectral
slope $\alpha$ is in the range $1.3-1.8$, with an average value $\alpha
= 1.51$ (see Fig.1).
To provide a frame of reference, it is convenient to consider two well known
cases: white noise has $\alpha = 0$ and Brownian motion has $\alpha = 2$.

\begin{figure}
\plotone{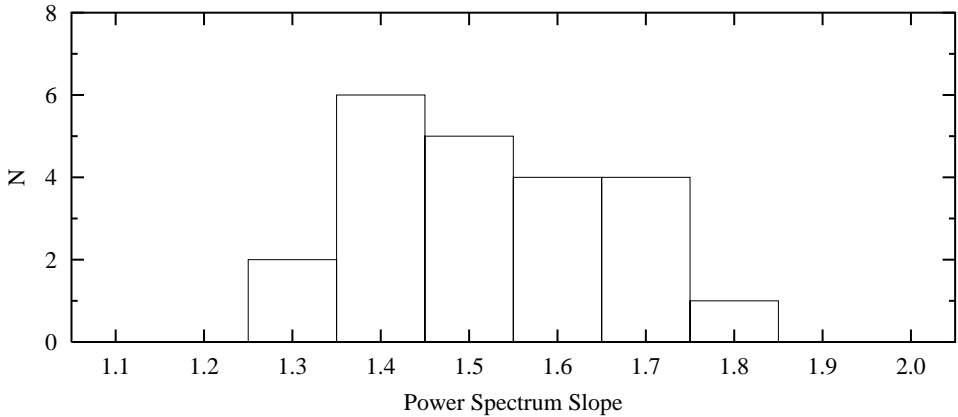}
\caption{Distribution of the PSD slopes ($\alpha$). The estimates are done 
using the SF and DACF, with a sample of blazars observed at 
the Perugia Astronomical Observatory.} 
\end{figure}

The point is that this kind of variability is observed in all the AGN class (see, e.g, 
the results of Webb \& Malkan 2000), and it does not seem to be exclusive of the
optical wavelengths: {\em AGNs seem to be characterised by variability similar 
to pink (or red) noise in a large range of frequencies}. The X-ray light curves have 
power spectra with slopes in the range $1$ to $2$, (see, e.g., Lawrance \& Papadakis 1993).
Radio light curves have power spectra with slopes around $2$ in time-scales
from days up to some years (see, e.g., Hufnagel \& Bregman 1992, Lainela
\& Valtaoja 1993). Moreover, we expect to see a similar behavior at higher energies too, 
because in the majority of theoretical models for AGNs there is a strictly correlation 
between $\gamma$-ray photons and low-frequency ones.
If this scenario will be confirmed also with future observations, then the
problem will be to identify the real time-scales and to understand the physical
reasons behind this kind of variability. For the moment, we can try to use
this observational evidence to improve our techniques in the research of
this class of objects during all-sky surveys.

\section{Application to the next high-energy missions}

EGRET detected 271 point sources (Hartman et al. 1999), but only 101 of them
were identified and, of them, 93 are AGNs. The next generation gamma-ray instruments
probably will increase this number to many thousands, but it is often crucial to
discriminate faint variable sources against the background noise.

\begin{figure}
\plottwo{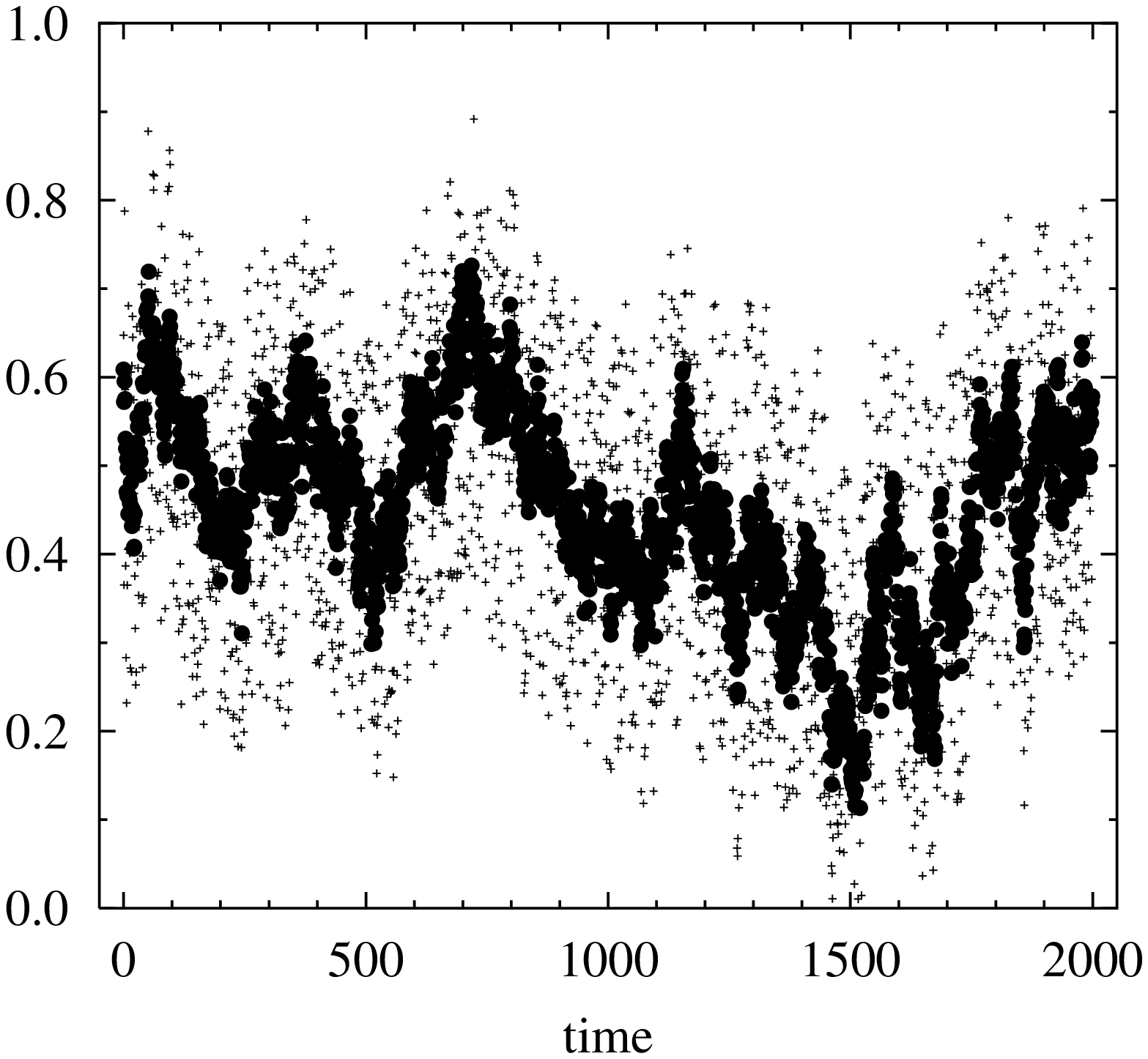}{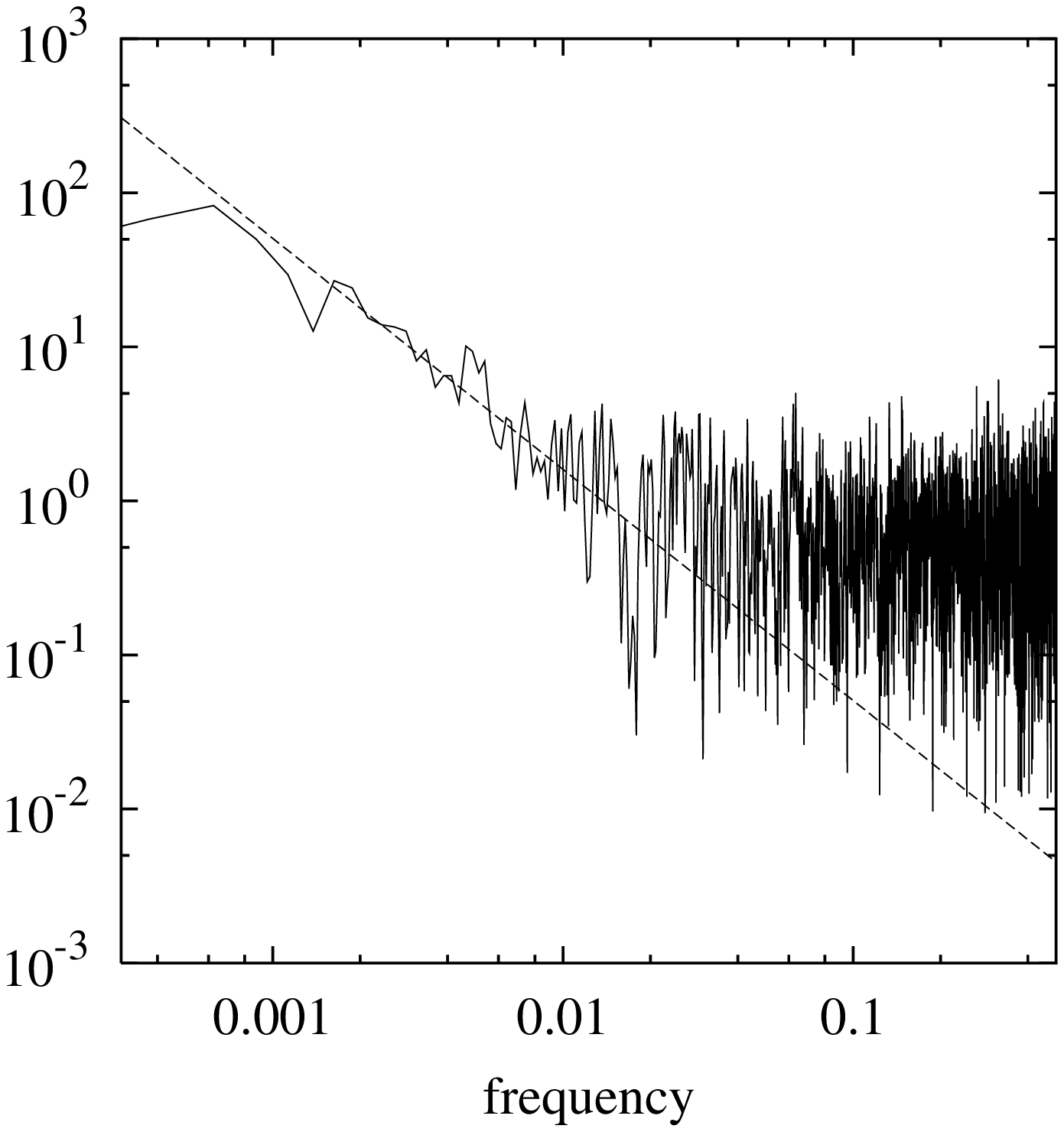}
\caption{The left panel shows the simulated ``identification'' of a variable
source (filled circles) from a noisy time-series (crosses). The signal is corrected
with the optimal (Wiener) filter, assuming that the PSD has a $f^{-1.5}$ behaviour
added to a white noise tail (right panel). Arbitrary units are used.}
\end{figure}

For this kind of problem it is generally used the likelihood analysis (Mattox et al., 
1996), that take into account the spatial and energetic distributions of the data.
Together with the more traditional techniques developed for a correct discovery
and identification of high-energy point sources, we believe that it would be useful
to use also the variability tools, for a correct inclusion of the time domain yet
during the first elaboration. In fact many astronomical sources (as AGNs, black 
hole X-ray binary systems, and also Gamma Ray Bursts) show variability characterised as
{\em red-noise} in a large scale of frequencies.

For example, Fig.2 shows the application of a traditional technique 
(the Wiener filter, see Press et al. 1992) for the ``identification'' of a 
simulated variable source with the assumption that the power spectrum has a 
$f^{-\alpha}$ behavior added to a white noise tail. With the original simulated 
signal it is quite difficult to verify the presence of a variable source respect to the
background noise (assumed as white noise), because the fluctuation is within the limit
of three standard deviations. The filtered signal puts in evidence the presence of a 
``possible'' point source, characterized by {\em red-noise} variability.

\section{Final remarks}
All the above mentioned statistical methods can be very useful for a rough
estimation of the physical phenomena,
but unfortunately they are not conclusive because the results are
always strongly affected by many external factors. For example, the Structure
Function seems to show a strong dependence from the more intense flares, while
the Discrete Correlation Function and the Detrended Fluctuation Analysis are
influenced by the large gaps in the light curves.

Moreover, it is not clear if the {\em red-noise} behavior is a typical 
feature of AGNs at the gamma-rays frequencies too, and what is the statistical
behavior of the background noise for the next generation high-energy detectors.

For all this reasons we are developing a Monte Carlo technique to
test all the more useful statistical descriptors, and to produce a reliable
quantification of the results so obtained.
In the same time, we intend to develop a neural network technique that
can be used to analyse the time series, to find the embedding dimension
of the characteristic attractor, and to perform well the frequency extraction 
in unevenly sampled signals. Finally, we are developing a light simulator for the
Large Area Telescope (LAT) onboard of the GLAST satellite, with the aim to test
the possible use of variability tools to discover and identify $\gamma$-rays
point sources.


\begin{references}
\reference Edelson, R. A. \& Krolik, J. H. 1988, ApJ, 333, 646
\reference Fiorucci, M., Tosti, G. \& Luciani, M. 1999, MSAIt, 70, 223
\reference Fiorucci, M. \& Tosti, G., 2000, in {\it Blazar monitoring towards the 3th millennium},
Raiteri C. \& Villata M. eds., 99
\reference Grassberger, P. \& Procaccia, I. 1983, Phys. Rev. Lett. 50, 346
\reference Hartman, R. C., Bertsch, D. L., Bloom, S. D., et al. 1999, ApJS, 123, 79
\reference Hufnagel, B. R. \& Bregman, J. N. 1992, ApJ, 386, 473
\reference Lainela, M. \& Valtaoja, E. 1993, ApJ, 416, 485
\reference Lawrence, A. \& Papadakis, I. 1993, ApJ, 414, 85
\reference Liszka, L. \& Holmstr\"om, M. 1999, A\&AS, 140, 125
\reference Lomb, N. R. 1976, Ap. Space Sci., 39, 447
\reference Mattox, J. R., Bertsch, D. L., Chiang, J., et al. 1996, ApJ, 461, 396
\reference Peng, C. K., Havlin, S., Stanley, H. E., et al. 1995, CHAOS, 5, 82 
\reference Press, W. H., et al. 1992, {\it Numerical Recipes: The Art of Scientific Computing}, 
Cambridge University Press
\reference Scargle, J. D. 1982, ApJ, 263, 835
\reference Tosti, G., Pascolini, S. \& Fiorucci, M. 1996, PASP, 108, 706
\reference Webb, W. \& Malkan, M. 2000, ApJ, 540, 652
\end{references}
\end{document}